

\magnification=\magstep1
\vsize=24truecm
\hsize=15.5truecm
\parskip=7pt
\baselineskip=16pt
\tolerance=5000
\font\grand=cmbx10 at 14.4truept

\def\\{{\hfil\break }}
\def\half{{1 \over 2}}
\def\det{{\rm det}}
\def\ts{\textstyle}

\def\z{\zeta}
\def\s{\sigma}
\def\a{\alpha}

\def\g{\gamma}
\def\d{\delta}
\def\m{\mu}
\def\b{\beta}

\def\blacksquare{\vrule height5pt depth0pt width5pt}
\def\ord{{\rm ord}}
\def\min{{\rm min}}
\def\no{{\rm N}}

\def\and{{\rm and}}
\def\mij{m_{ij}}
\def\zp{\zeta_{p^k}}
\def\zq{\zeta_{q^l}}

\pageno=0
\def\folio{
\ifnum\pageno<1 \footline{\hfil} \else\number\pageno \fi}

\rightline{DIAS--STP--92--45}

\vskip 3truecm

\centerline{\grand
Automorphisms of the affine SU(3) fusion rules}

\vskip 3truecm

\centerline{Philippe Ruelle \footnote{$^*$}{e--mail address:
ruelle@stp.dias.ie}}
\vskip 1truecm
\centerline{\it Dublin Institute for Advanced Studies}
\centerline{\it 10 Burlington Road}
\centerline{\it Dublin 4, Ireland}

\vskip 3truecm

\centerline{\bf Abstract}
\medskip
\leftskip 1.2truecm
\rightskip 1.2truecm
\noindent
We classify the automorphisms of the (chiral) level--$k$ affine
$SU(3)$ fusion rules, for any value of $k$, by looking for
all permutations that commute with the modular matrices $S$ and $T$.
This can be done by using the arithmetic of the cyclotomic extensions
where the problem is naturally posed. When $k$ is divisible by 3, the
automorphism group ($ \sim Z_2$) is generated by the charge
conjugation $C$. If $k$ is not divisible by 3, the automorphism
group ($\sim Z_2 \times Z_2$) is generated by $C$ and the
Altsch\"uler--Lacki--Zaugg automorphism. Although the combinatorial
analysis can become more involved, the techniques used here
for $SU(3)$ can be applied to other algebras.

\leftskip=0cm
\rightskip=0cm

\vfill\eject

\noindent
{\bf 1. Introduction.}

\noindent
Modular invariance has received much attention over the past six
years, as it proved to play a key r\^ole in the classification of
2d conformal field theories [1]. For a left--right symmetric
theory, the basic problem is to classify the
modular invariant partition functions of the form
$$Z(\tau ^*,\tau) = \sum_{i,j} \; \chi_i^*(\tau) \, N_{ij} \,
\chi_j(\tau),
\eqno(1.1) $$
where the $\chi_i(\tau)$, possibly in infinite number, are the
irreducible characters of the chiral symmetry algebra occurring
in that theory. The matrix $N$ in (1.1) must have non--negative integer
entries and must be normalized by requiring $N_{00}=1$, where
$\chi_0$ denotes the character of the representation which contains
the (chiral) vacuum. The characters carry a representation of the
modular group:
$$\chi_i(\tau +1) = \sum_j \;T_{ij}\,\chi_j(\tau), \quad {\rm and}
\quad \chi_i({\ts {-1 \over \tau}}) = \sum_j \;S_{ij}\,\chi_j(\tau).
\eqno(1.2) $$
That $Z(\tau ^*,\tau)$ is modular invariant forces $N$ to satisfy
$$T^\dagger \,N\,T=N \quad {\rm and} \quad S^\dagger \,N\,S=N.
\eqno(1.3) $$
When the modular matrices $S$ and $T$ are unitary, the conditions
(1.3) are equivalent to $N$ being in their commutant:
$[T,N]=[S,N]=0$.

The above conditions on the matrix $N$ prove to be extremely
restrictive. A general analysis was carried out by Moore and
Seiberg [2]. Their result is that, for a given theory, the
matrices $N$ which satisfy all the conditions must be permutation
matrices, or else they are such once the symmetry has been adequately
extended. Moreover, as follows from the Verlinde's formula [3],
these permutations are automorphisms of the fusion coefficients of
the original or the extended theory respectively.

For only a small class of theories has the classification been
completed. Examples (almost all related to each other)
include theories with an affine $SU(2)$ symmetry
[4], the (unitary and non--unitary) Virasoro minimal models [4],
supersymmetric minimal models [5] and parafermionic theories [6].
The only non--rational theories for which a classification is
known are the conformal theories with $c=1$ [7].

Among the rational theories, those with an affine Lie symmetry play
a central r\^ole as they are thought to be the building blocks to
construct all the others. At present, the complete classification is
known only for theories with an $\widehat{SU(2)}$ symmetry [4],
although partial results exist for $\widehat{SU(3)}$ [8,9].

The purpose here is to study the modular invariant
partition functions of theories possessing a symmetry not larger than
an (untwisted) affine Lie symmetry. In other words, we will be looking
for permutations $N$ commuting with the matrices $S$ and $T$
describing the modular transformations of the characters of
Kac--Moody algebras. Even so, the problem is still considerable if
one wishes to classify them all. Here we restrict ourselves to
the $\widehat{SU(3)}$ algebra, which is the simplest case still open
and yet, which offers generic features of other simple algebras.

\medskip
The integrable representations of the $\widehat{SU(3)}_k$
Kac--Moody algebra are
in correspondence with the $SU(3)$ strictly dominant weights $p$ in
the alc\^ove $B_n=\{p=(a,b) \;:\; a,b \geq 1 \; {\rm and} \;
a+b \leq n-1\}$, where we set the height $n=k+3$ [10]. Their total
number is $(n-1)(n-2) \over 2$. The representation labelled by
$p=(1,1)$ contains the vacuum of the Fock space where the algebra is
being represented. We denote by $\chi_p(\tau)$ the corresponding
(restricted) characters. As functions of $\tau$, we have
$\chi_p(\tau)=\chi_{p'}(\tau)$ if and only if $p'=Cp$ where $C$ is
the charge conjugation acting by $C(a,b)=(b,a)$.

The modular matrices, unitary in this case, have the following
expressions. For $p=(a,b)$ and $p'=(c,d)$, the $T$ matrix reads
$$T_{p,p'} = \exp\Big[2i\pi \big({p^2 \over 2n}-{1 \over 3}\big)\Big]
\,\delta_{p,p'} =
\z_{3n}^{a^2+ab+b^2-n} \; \delta_{a,c}\,\delta_{b,d},
\eqno(1.4a)$$
while the $S$ matrix is more complicated
$$\eqalign{
S_{p,p'} & = {-i \over \sqrt{3}n} \,\sum_{w \in W} \,
(\det\,w) \,\exp\Big[2i\pi {p \cdot w(p') \over n} \Big], \cr
& = {-i \over \sqrt{3}n} \big\{
\zeta_{3n}^{(2a+b)c+(a+2b)d} + \zeta_{3n}^{-(a+2b)c+(a-b)d} +
\zeta_{3n}^{-(a-b)c-(2a+b)d} \cr
& \qquad \qquad -\zeta_{3n}^{(2a+b)c+(a-b)d}-\zeta_{3n}^{-(a-b)c+(a+2b)d}
-\zeta_{3n}^{-(a+2b)c-(2a+b)d} \big\}.}
\eqno(1.4b) $$
Here $\z_{3n}=\exp{({2i\pi \over 3n})}$ and $W=S_3$ is the Weyl group
of $SU(3)$.

In the following, we classify, for all heights $n$, the permutation
matrices $N_{p,p'}=\delta_{p',\s(p)}$ which commute with the matrices $S$
and $T$ of (1.4), thereby classifying the partition functions
of the form
$$Z(\tau,\tau^*)=\sum_{p \in B_n} \, [\chi_p(\tau)]^*
\,[\chi_{\s(p)}(\tau)].
\eqno(1.5) $$

Since the permutations $\s$ are also automorphisms of the fusion
rules, we could try to determine them directly from the fusion
coefficients. This is indeed possible for $SU(3)$, by using their
explicit expressions, obtained recently in [11]. It would however
definitely confine us to $SU(3)$ since the fusion coefficients for
higher ranks are not known. Instead, the approach we follow here,
although applied to $SU(3)$, does not confine us to this particular
case. We emphasize that we will not use any peculiar feature of
$SU(3)$ that is not immediately available in other algebras. Our
analysis can therefore be carried out in other cases as well. Another
advantage of looking at the $S$ matrix elements is that our proof can
be useful to classify the automorphisms of the extensions defined by
the complementary invariants of [12]. Indeed for those extensions,
most of the extended $S$ matrix is the same as the non--extended one,
up to numerical factors.

Finally we should mention that modular invariants of the kind we are
interested in here are already known. Whenever the KM algebra has
outer automorphisms, Altsch\"uler, Lacki and Zaugg have shown that
one can construct a whole class of invariants (also called
complementary) [13]. Whether their
invariants are exhaustive is generally an open problem, though
the invariants found in [14] for $G_2$ and $F_4$ show that they are
not exhaustive in those cases at least. For
$\widehat{SU(3)}$, we will show that they are complete.

\vskip 0.5truecm \noindent
{\bf 2. The classification.}

\noindent
The best part of this article will be devoted to the proof of the
following necessary condition for $\s$ to commute with $S$.
\vskip 0.3truecm \noindent
{\bf Theorem.} Let $p=(a,b)$ and $p'=\s(p)=(c,d)$ two weights in
the alc\^ove related by an automorphism $\s$.
Then, modulo $n$, $(c,d,-c-d)$ is a permutation of $(a,b,-a-b)$.

\medskip \noindent
The Theorem can be proved by only requiring that $\s$ commutes with
$S$, although for simplicity, we will make use of a stronger
condition. Its proof is contained in Sections 4 and 5. For the moment
we show that the classification follows from it.

Since the weight $p'=\s(a,b)$ must belong to the alc\^ove, the
six values quoted in the Theorem are
$$\s(a,b) = \cases{
(a,b),\, (n-a-b,a),\, (b,n-a-b), \cr \noalign{\smallskip}
(b,a),\, (a,n-a-b),\, (n-a-b,b). \cr}
\eqno(2.1) $$
The last three values are the charge conjugated of the first
three. We first ignore the action of the charge conjugation $C$,
therefore focusing on the coset of the automorphism group by $C$.

\smallskip
The first three weights in (2.1) are the images of $(a,b)$ under
the outer automorphisms of $\widehat{SU(3)}$, generated by
$\m(a,b)=(n-a-b,a)$, $\m^3=1$. One readily
checks that for any pair of weights $p,p'$ in the alc\^ove,
$$ S_{\m^k(p),p'} = {\rm e}^{-{2i\pi kt(p') \over 3}} \, S_{p,p'},
\eqno(2.2) $$
where $t(p')=c-d \bmod 3$ is the triality of the weight
$p'=(c,d)$.

So the Theorem says that the pointwise action of an automorphism of
the fusion rules must be an outer automorphism of the KM algebra, up
to the charge conjugation. The problem is to
define $\s$ on the whole of $B_n$ in such a way that it still
commutes with $S$. On the other hand,
$\s$ must also commute with $T$, which implies, from (1.4$a$), that
the norms of $p$ and $\s(p)$ must be equal modulo $2n$. From
$$ (\m^k(p))^2 = p^2 + {\ts {2n \over 3}}[n - k t(p)] \bmod 2n
\qquad {\rm for\ }k \neq 0,
\eqno(2.3) $$
we obtain the following possibilities, depending on the residue of
$n$ modulo 3 and the triality of $p$:
$$\eqalign{
n=0 \bmod 3 \quad : \quad & \s(p)=\m^k(p) \quad \hbox{if $t(p)=0$}, \cr
& \s(p)=p, \quad \hbox{if $t(p) \neq 0$}, \cr
n \neq 0 \bmod 3 \quad : \quad & \s(p)=p \quad {\rm or}
\quad \m^{n t(p)}(p). \cr}
\eqno(2.4) $$
We now impose the commutation of $\s$ with $S$, which reads
$$S_{\s(p),p'} = S_{p,\s^{-1}(p')} \qquad {\rm for\ all\ }
p,p' \in B_n.
\eqno(2.5) $$

If $n=0 \bmod 3$, for any fixed root $p$ of zero triality, we choose
a weight $p'$ of non--zero triality such that $S_{p,p'} \neq 0$. (This
is always possible unless $p=({n \over 3},{n \over 3})$, but then
$\mu (p)=p$ anyway.) We obtain from (2.2) and (2.4)
$$S_{\s(p),p'} = S_{\m^k(p),p'} = {\rm e}^{-{2i \pi k t(p') \over 3}} \,
S_{p,p'} = S_{p,\s^{-1}(p')} = S_{p,p'}.
\eqno(2.6) $$
Equation (2.6) implies $k=0$, so that none of the weights in $B_n$,
whatever its triality, can undergo a non--trivial transformation
$\s$ (up to $C$).

For $n \neq 0 \bmod 3$, we take $p=(n-2,1)$ and an arbitrary weight
$p'$, both of non--zero triality, and
prove that if $p$ undergoes a non--trivial transformation, then $p'$
has to do the same. Suppose the contrary, namely
$\s(p)=\m^{nt(p)}(p)$ and $\s(p')=p'$. We have from (2.2)
$$S_{\s(p),p'} = S_{\m^{nt(p)}(p),p'} = {\rm e}^{-{2i \pi nt(p)t(p')
\over 3}} \, S_{p,p'} = S_{p,\s^{-1}(p')} = S_{p,p'}.
\eqno(2.7) $$
The matrix element $S_{p,p'}=S_{(n-2,1),p'}$ is never zero for any
$p'$, so that (2.7) is a contradiction since $nt(p)t(p')$ is not zero
modulo 3. Thus the transformation $\m^{nt(\cdot)}(\cdot)$
acts on all the weights of $B_n$ or on none of them.

\smallskip
We have proved that, up to the charge conjugation $C$, there is no
automorphism if $n=0 \bmod 3$, and there is a single one if $n \neq
0 \bmod 3$, acting by $\s(p)=\m^{nt(p)}(p)$. This non--trivial
automorphism is a permutation of order 2.

Finally we show that the charge conjugation must act in the same way
on all the weights in $B_n$ if it is to commute with $S$. From
$$S_{C(p),p'} = S^*_{p,p'},
\eqno(2.8) $$
we obtain that, if $p$ is transformed by $C$ while $p'$ is kept
fixed, $S_{p,p'}$ must be real.
However the equation (2.2) implies
$$S_{\m(1,1),p'} = S_{(n-2,1),p'} = {\rm e}^{-{2i \pi t(p') \over
3}} \, S_{(1,1),p'}.
\eqno(2.9) $$
Since the matrix element $S_{(1,1),p'}$ is real and strictly
positive for any $p'$, it follows that $S_{(n-2,1),p'}$
has a non--zero imaginary part for every $p'$ with a non--zero
triality. Thus if a weight $p'$ is conjugated, then $(n-2,1)$ must
also be conjugated, and in turn that means that every weight
has to be conjugated. Therefore, $C$ acts on
all the weights of non--zero triality or on none of them. To settle the
question for the roots, we go back to the definition
of $\s$ as an automorphism of the fusion rules.

It is straightforward to compute the fusion rule of the fundamental
representation of $SU(3)$ with any other representation. The result
is (in terms of the shifted weights)
$$(2,1) \, * \, (a,b) = (a+1,b) \, + \, (a-1,b+1) \, + \, (a,b-1),
\eqno(2.10) $$
where however, on the right-hand side, a representation must be
omitted if one of its Dynkin label is zero or if the sum of its
Dynkin labels is equal to $n$. If we take a root for $(a,b)$,
all the other representations entering (2.10) have non--zero triality.
This shows that if none of the weights undergoes the $C$ transformation,
none of the roots can either and conversely, if the fusion rules (2.10)
are to be kept invariant. Therefore the charge conjugation $C$ is an
automorphism of the fusion rules if and only if it transforms
uniformly all the weights and roots of the alc\^ove.

The proof is complete. We note that for $n=4$ and 5, the
actions of $\m^{nt(\cdot)}(\cdot)$ and $C$ are identical.
We have the

\vskip 0.5truecm \noindent
{\bf Proposition.} {\sl The automorphism group of the fusion rules
of $\widehat{SU(3)}_k$ is generated by $C$ if $n=k+3$ is divisible by 3
or if $n=4$ or $5$, and is generated by $C$ and $\m^{nt(\cdot)}(\cdot)$
when $n \geq 7$ is not divisible by 3. The group structure
is $Z_2$ and $Z_2 \times Z_2$ respectively.}

\bigskip \noindent
As a direct consequence, there exist respectively two or four modular
invariant partition functions originating from automorphisms of the
fusion rules. They are the only ones if the $\widehat{SU(3)}$
symmetry is not extended.

\vskip 0.5truecm \noindent
{\bf 3. Preliminaries.}

\noindent
The proof of the Theorem of Section 2 extensively uses the arithmetic
of cyclotomic fields. A useful reference on this matter is the book by
Washington [15].

Let $\z_n$ be a primitive $n$--th root of unity, for an arbitrary
integer $n$, and let $Q(\z_n)$ denote the corresponding cyclotomic
extension, of degree $\varphi(n)$ over the rationals. Its
Galois group, noted ${\rm Gal}(Q(\z_n)/Q)$, is isomorphic to
$Z_n^*$ (the group of integers invertible modulo $n$) and
transforms $\z_n$ into $\z_n^\a$ for $\a$ coprime with $n$.

If $p^l$ divides $n$, $Q(\z_n)$ is an algebraic extension of
$Q(\z_{n/p^l})$, of relative degree $p^l$ or $p^{l-1}(p-1)$
according to whether $p$ does
or does not divide ${n \over {p^l}}$. In each case, the extension can be
defined by the irreducible polynomial $X^{p^l} - \z_{n/p^l} = 0$ and
$\Phi_{p^l}(X) = 0$ respectively, where $\Phi_m(X)$ denotes the $m$--th
cyclotomic polynomial.
If $k={\rm ord}_pn$ ($p^k$ is the largest power of $p$ dividing
$n$), the Galois group of the relative extension is:
$$\eqalign{
{\rm Gal}(Q(\z_n)/Q(\z_{n/p^l})) & =
\{\s_\a(\z_n)=\z_n^\a \;:\; \a=1 \bmod {n \over p^l} {\rm \ and\ }
(\a,n)=1 \} \cr
& \sim Z_{p^l} \;(l<k) \quad {\rm or} \quad Z_{p^l}^* \;(l=k). \cr}
\eqno(3.1) $$

For any $z$ in $Q(\z_n)$, one defines its norm (over $Q$) by taking
the product of all its Galois conjugates: $\no_{Q(\z_n)/Q}(z)=
\prod_{\s \in {\rm Gal}(Q(\z_n)/Q)}\,\s(z)$.
For $d$ a divisor of $n$ and $x$ an integer coprime with $n \over d$,
one obtains
$$\no_{Q(\z_n)/Q}(1 - \z_n^{dx}) = \cases{
1 & if two different primes divide ${n \over d}$, \cr
p^{{\varphi(n)} \over {\varphi(n/d)}} & if $p$ is the only prime
dividing ${n \over d}$. \cr}
\eqno(3.2) $$

We also note the useful polynomial identity
$$\prod_{j=1}^{m} \, (1-X\,\z_m^j) = 1-X^{m}.
\eqno(3.3) $$

\medskip
In the maximal real sub--field $Q(\z_n+\z_n^{-1})$, the following
subset of cyclotomic units will have some importance.
Let $n=p^k$ be a prime power. These units are defined by
$$\xi_a = \z_n^{(1-a)/2} \; {{1 - \z_n^a} \over {1 - \z_n}}, \qquad
1 < a < {n \over 2},\quad (a,n)=1.
\eqno(3.4) $$
All the $\xi_a$ are real and their number is equal
to $r=\half \varphi(n)-1$, although the $\xi_a$
can be defined for any $a \in Z_n^*$ and satisfy
$\xi_a + \xi_{-a} = 0$. In particular, $\xi_1=1$ and $\xi_{-1}=-1$.
The most useful property of the units $\xi_a$ is that
they are multiplicatively independent in $Q(\z_n+\z_n^{-1})$.
It means that the existence of the relation
$$ \xi_{a_1}^{t_1} \, \xi_{a_2}^{t_2} \, \ldots \xi_{a_r}^{t_r} =
(-1)^{t_0}, \qquad t_i \in Z,
\eqno(3.5) $$
requires $t_1=t_2=\ldots=t_r=0$ and $t_0$ be even.

We will also need (additive) independence properties among the
roots of unity. Let again $n=p^k$. A complete set of relations is
given by
$$\z_n^r \, \big( 1 + \z_n^{p^{k-1}} + \z_n^{2p^{k-1}} + \ldots
+ \z_n^{(p-1)p^{k-1}} \big)=0, \quad 0 \leq r \leq p^{k-1}-1.
\eqno(3.6r) $$
Note that each of the $n$ powers of $\z_n$ appears in one and only
one relation. This implies that if a set of powers $\z_n^{a_i}$ is
not linearly independent, the equation (3.6$r$) for some $r$ must
hold among $p$ of them. In particular, any set of $N<p$ different
powers is linearly independent.

The related independence problem for $n$ not a prime power can
be reduced to the above case by using the fact that $Q(\z_{mn})$ is
the product of $Q(\z_m)$ and $Q(\z_n)$ if $m$ and $n$ are coprime:
one can choose a basis of $Q(\z_{mn})$ which is the product of the
bases of $Q(\z_m)$ and $Q(\z_n)$. This property implies
that if a set of powers $\z_n^{a_i} \in Q(\z_n)$ are linearly
independent over $Q$, they are also linearly independent over
$Q(\z_m)$ provided $(n,m)=1$.

\smallskip
Our starting point to prove the Theorem is
the expression (1.4$b$) for the matrix
elements of $S$. When one of the indices is a \lq diagonal' root
$(l,l)$, the expression simplifies to become (from now on, we omit the
prefactor $-i \over \sqrt{3}n$)
$$S_{(l,l),(a,b)} = \z_n^{la+lb} + \z_n^{-la} + \z_n^{-lb} - {\rm c.c.}
\eqno(3.7) $$
This is an additive form of $S_{(l,l),(a,b)}$. In view of the
independence property of the units (3.4), the following
multiplicative form is equally useful. It is obtained by using the
expression for the denominator of the Weyl character formula
$$S_{(l,l),(a,b)} = (1-\z_n^{la}) (1-\z_n^{lb}) (1-\z_n^{-la-lb}).
\eqno(3.8) $$
Let us recall the generalization of (3.8) to any simple algebra
$\hat G_k$. We set $n=k+h$ with $h$ the dual Coxeter number of $G$.
When $p=l\rho$ is a weight proportional to $\rho$, half
the sum of the positive roots, the Weyl formula recasts the matrix
element $S_{p,p'}$ into (up to an irrelevant prefactor)
$$S_{p,p'} = S_{(l,l,\ldots,l),p'} =
\prod_{{\rm positive\ roots\ }\a} \,\,\z_{n}^{l \a \cdot p'/2}\;
(1-\z_{n}^{-l \a \cdot p'}),
\eqno(3.9) $$
On account of the definition (3.4), $S_{l\rho,p'}$ can be expressed
as a product of units $\xi_a$, up to an overall power of $(1-\z_n)$ and
$\z_n$. This formula is the main tool of Section 4.
(Note that if $G$ is not simply--laced, the numbers
$\a \cdot p'$ may not be integers.)

\smallskip
We also recall the arithmetical symmetry that the commutant of $S$
and $T$ was recently shown to possess [9]. Let $N$ be a matrix commuting
with $S$ and $T$. ($N$ can have complex entries.) One
defines on the pairs of $B_n \times B_n$ the following action of the
group $Z_{3n}^*$. For any $\nu \in Z_{3n}^*$, it is defined
by $M_\nu \;:\; (p,p') \longrightarrow (p_\nu,p'_\nu)$ where $p_\nu
\in B_n$ is the image by an affine Weyl transformation $w_\nu$ of
the weight $\nu p$. The symmetry was the statement that under this
action, the coefficients $N_{p,p'}$ of $N$ satisfy
$$N_{p,p'} = (\det \,w_\nu)(\det \,w'_\nu) \; N_{p_\nu,p'_\nu}.
\eqno(3.10) $$
In particular it was noted that $M_{-1}(p)=Cp$ is the charge
conjugation, implying $N_{p,p'}=N_{Cp,Cp'}$ for any $p,p'$. As
a consequence, if $N$ is to be a permutation
matrix, a diagonal root can only be permuted with another
diagonal root: $p=Cp$ and $N_{p,p'} \neq 0$ imply $p'=Cp'$. In the
following, we use this mild property in the only purpose to
simplify the proofs. The Theorem can be proved without using it.
(In general, one finds $M_{-1}(p)=Cp$ for
$G=SU(N),\,SO(4N+2)$ and $E_6$, while $M_{-1}(p)=p$ is the identity
in all other cases, $-1$ being a Weyl transformation.)

\medskip
The following two sections contain the proof itself of the Theorem. We
will exclusively use the matrix elements $S_{(l,l),p}$ in the form
(3.7) and (3.8). Section 4 is essentially multiplicative while
Section 5 is definitely additive.

\vskip 0.5truecm \noindent
{\bf 4. A local version of the Theorem.}

\noindent
Throughout this section and the next one, we let $n=\prod_1^s p_i^{k_i}$
be the prime decomposition of $n$, so that $k_i=\ord_{p_i}n$.

In this section, we prove that the Theorem is (almost) true if we
replace the congruence modulo $n$ by a congruence modulo
$p_i^{k_i}$, for any $i$ (Corollary 1).
We set $(c,d)=\s(a,b)$. They must satisfy $a,b,a+b,c,d,c+d \neq 0
\bmod n$ to be in the alc\^ove $B_n$.
The core of the analysis is contained in the following lemma, concerned
with the solutions of the following two equations
$$\eqalignno{
& (1-\z_n^a)(1-\z_n^b)(1-\z_n^{-a-b}) = (1-\z_n^c)(1-\z_n^d)
(1-\z_n^{-c-d}), &(4.1) \cr
& (1-\z_{p^{k}}^a)(1-\z_{p^{k}}^b)(1-\z_{p^{k}}^{-a-b}) =
(1-\z_{p^{k}}^c) (1-\z_{p^{k}}^d) (1-\z_{p^{k}}^{-c-d}).
&(4.2) \cr}
$$
Equation (4.1) expresses the fact that $[S,\s]_{(1,1),(a,b)}=0$, as
follows from (2.5) and (3.8), and the invariance of (1,1) under any
automorphism. Likewise, (4.2) is $[S,\s]_{({n \over
p^k},{n \over p^k}),(a,b)}=0$ if $({n \over
p^k},{n \over p^k})$ is known to be invariant under $\s$.

\smallskip
Let us define $l_x=\ord_p x$ for
$x=a,b,a+b,c,d,c+d$. We note that within each triplet
$(l_a,l_b,l_{a+b})$ or $(l_c,l_d,l_{c+d})$, two numbers must be
equal and furthermore, these two are smaller or equal to the third
one, because of
$$l_{a+b} \geq \min(l_a,l_b), \qquad l_{c+d} \geq \min(l_c,l_d),
\eqno(4.3) $$
where the equalities hold if $l_a \neq l_b$ or $l_c \neq l_d$.

\vskip 1truecm \noindent
{\bf Lemma 1.}
Let $a,b,c,d$ be integers such that $a,b,a+b,c,d,c+d \neq 0 \bmod
n$ satisfy the equations (4.1) and (4.2), where $k=\ord_p n$ .
Then either $(c,d,-c-d)$ is a permutation of $(a,b,-a-b)$ mod $p^k$,
or else we must have (up to permutations of $a,b,a+b$ or of
$c,d,c+d$):\\
\hangindent=1cm \hangafter=4
$p=2,3 \;:\; l_a=l_b=l_{a+b}=l_d=k \quad {\rm and} \quad
l_c=l_{c+d}=k-1,$ \hfill (4.4$a$)\\
$p=2 \;:\; l_a=l_b=l_{a+b}=l_d=k \quad {\rm and} \quad
l_c=l_{c+d}=k-2,$ \hfill (4.4$b$)\\
$p=2,3 \;:\; l_c=l_d=l_{c+d}=l_b=k \quad {\rm and} \quad
l_a=l_{a+b}=k-1,$ \hfill (4.4$c$)\\
$p=2 \;:\; l_c=l_d=l_{c+d}=l_b=k \quad {\rm and} \quad
l_a=l_{a+b}=k-2,$ \hfill (4.4$d$)\\
$p=2 \;:\; l_a=l_{a+b}=k-1,\;\; l_b=k, \quad
{\rm and} \quad l_c=l_{c+d}=k-2,\;\; l_d=k,$ \hfill (4.4$e$)\\
$p=2 \;:\; l_c=l_{c+d}=k-1,\;\; l_d=k, \quad
{\rm and} \quad l_a=l_{a+b}=k-2,\;\; l_b=k.$ \hfill (4.4$f$)

\noindent {\sl Proof.}
Due to the symmetry of the problem, we need to consider only four
cases: $l_a=l_b=l_{a+b}<k$ (case 1), $l_a=l_{a+b}<l_b<k$ (case 2),
$l_a=l_b=l_{a+b}=k$ (case 3) and finally $l_a=l_{a+b}<l_b=k$ (case 4).

\medskip \noindent
$\underline{\hbox{Case 1.} \quad l_a=l_b=l_{a+b}<k.}$\\
Let $l=l_a$. Without loss of generality, we can assume $l_c=l_{c+d}
\leq l_d<k$. (None of $l_c,l_d,l_{c+d}$ can be equal to $k$ since
the left--hand side of (4.2) is not zero.)
Taking the norm $\no_{Q(\z_{p^k})/Q}$ of (4.2), we obtain from (3.2)
$$3p^l = 2p^{l_c} + p^{l_d}.
\eqno(4.5) $$
If $l_c$ and $l_d$
are not both equal to $l$, one is smaller and the other is bigger
than $l$, that is $l_c < l < l_d$. Then (4.5)
yields $2p^{l_c}=0 \bmod p^l$, a contradiction unless $p=2$.
However $p=2$ is already
excluded from the very start, because it is not compatible with
$l_a=l_b=l_{a+b}$.

Hence $l_c=l_d=l_{c+d}=l$.
For $a=\a p^l,\,b=\b p^l,\,c=\g p^l,\,d=\d p^l$ with
$\a,\b,\g,\d,\a+\b,\g+\d$ coprime with $p$, (4.2) reads
$$(1-\z^{\a}) (1-\z^{\b}) (1-\z^{-\a-\b}) =
(1-\z^\g) (1-\z^{\d}) (1-\z^{-\g-\d}), \qquad \z=\z_{p^{k-l}}.
\eqno(4.6) $$
Dividing (4.6) by $(1-\z)^3$, we get $\xi_\a \xi_\b
\xi_{-\a-\b} \xi_\g^{-1} \xi_\d^{-1} \xi_{-\g-\d}^{-1} = 1$ from
(3.4). The independence
property of the $\xi$'s implies that $(\a,\b,-\a-\b)$ is a
permutation of $(\g,\d,-\g-\d) \bmod p^{k-l}$ and
therefore $(a,b,-a-b)$ is a permutation of $(c,d,-c-d)
\bmod p^k$, as required.

\medskip \noindent
$\underline{\hbox{Case 2.} \quad l_a=l_{a+b} < l_b < k.}$\\
Again we assume $l_c=l_{c+d} \leq l_d < k$. Now the norm from
$Q(\z_{p^k})$ to $Q$ of (4.2) yields
$$ 2p^{l_a} + p^{l_b} = 2p^{l_c} + p^{l_d}.
\eqno(4.7) $$
We cannot have $l_c=l_d=l_{c+d}<k$ because, from the Case 1,
it would imply $l_a=l_b=l_{a+b}$. So $l_c=l_{c+d} <
l_d < k$.

Assume first $l_d>l_b$. We obtain from (4.7)
$p^{l_c} = 0 \bmod p^{l_a}$ and
$2^{l_c+1} = 0 \bmod 2^{l_a+1}$ for $p \neq 2$ and $p=2$
respectively, implying $l_c \geq l_a$. Since (4.7) has no solution
for $l_d > l_b$ and $l_c > l_a$, we must have $l_c=l_a$, a
contradiction since it implies $l_d=l_b$.
We obtain the same contradiction if we assume $l_d < l_b$, by
exchanging the two triplets $(a,b,a+b)$ and $(c,d,c+d)$. Therefore
$l_d=l_b$ and $l_c=l_a$.

\smallskip
Setting $a=\a p^{l_a},\,b=\b p^{l_b},\,c=\g p^{l_a}$, and $d=\d
p^{l_b}$ with $\a,\b,\g,\d$ coprime with $p$, (4.2) becomes for
$\z=\z_{p^{k-l_a}}$
$$(1-\z^{\a}) (1-\z^{\b p^{l_b-l_a}}) (1-\z^{-\a-\b p^{l_b-l_a}}) =
(1-\z^{\g}) (1-\z^{\d p^{l_b-l_a}}) (1-\z^{-\g-\d p^{l_b-l_a}}).
\eqno(4.8) $$
Using (3.3) twice with $X=\z^\b$ or $\z^\d$ and $m=p^{l_b-l_a}$,
(4.8) can be recast into
$$\eqalign{
(1-\z^{\a}) \, (1-\z^{-\a-\b p^{l_b-l_a}}) \, &
\prod_{j=1}^{p^{l_b-l_a}} (1-\z^{\b+jp^{k-l_b}}) = \cr
& (1-\z^{\g}) \, (1-\z^{-\g-\d p^{l_b-l_a}}) \,
\prod_{j=1}^{p^{l_b-l_a}} (1-\z^{\d+jp^{k-l_b}}). \cr}
\eqno(4.9) $$
Dividing (4.9) by $(1-\z)^{2+p^{l_b-l_a}}$, we obtain
$$\xi_\a \, \xi_{-\a-\b p^{l_b-l_a}} \,
\prod_{j=1}^{p^{l_b-l_a}} \xi_{\b+jp^{k-l_b}} \, =
\xi_\g \, \xi_{-\g-\d p^{l_b-l_a}} \,
\prod_{j=1}^{p^{l_b-l_a}} \xi_{\d+jp^{k-l_b}} \,.
\eqno(4.10) $$
The sub--indices of the $\xi$'s are now all coprime with
$p^{k-l_a}$, so we can use their independence to obtain
$\g=\a \; {\rm or} \;
-\a-\b p^{l_b-l_a} \bmod p^{k-l_a}$ and
$\d=\b \bmod p^{k-l_b}$, or equivalently $(c,d)=(a,b)$ or
$(-a-b,b) \bmod p^k$. Restoring the symmetry, we have
that $(c,d,-c-d)$ is a permutation of $(a,b,-a-b)$ modulo $p^k$.

\medskip \noindent
$\underline{\hbox{Case 3.} \quad l_a=l_b=l_{a+b}=k.}$\\
Equation (4.2) shows that one of $c,d,c+d$ must be zero mod $p^k$
(since the left--hand side is zero). Suppose $d$ is the one and $l_c =
l_{c+d} \leq l_d = k$. We want to prove $l_c=l_{c+d}=k$ as well.

If $l_c < k$, {\sl i.e.} $c \neq 0 \bmod p^k$, every $\s_\a \neq 1$
in ${\rm Gal}(\z_n/\z_{n/p^k})/{\rm
Gal}(\z_n/\z_{n/p^{l_c}}) \sim Z^*_{p^{k-l_c}}$ is such that
$\s_\a(\z_n^c) \neq \z_n^c$. In other words, $\s_\a$ leaves $\z_n^a,
\z_n^b$ and $\z_n^d$ invariant, but not $\z_n^c$. Acting with
$\s_\a$ on (4.1) and comparing back with (4.1) yields
$$\z_n^c - \z_n^{\a c} = -\z_n^{-d} \, (\z_n^{-c} - \z_n^{-\a c}) =
\z_n^{-d-(\a+1)c} \, (\z_n^c - \z_n^{\a c}).
\eqno(4.11) $$
Equation (4.11) implies $d+(\a+1)c=0 \bmod n$. If we write
$c=c_1 p^{k} + \g p^{l_c} {n \over {p^{k}}}$, then
$\a c=c_1 p^{k} + j \g p^{l_c} {n \over {p^{k}}}$ for $j \neq 1$ in
$Z_{p^{k-l_c}}^*$ (see (3.1)). The condition $d+(\a+1)c=0 \bmod n$
implies $(1+\a)c = 0 \bmod p^{k}$, or
$$1+j=0 \bmod p^{k-l_c}.
\eqno(4.12) $$
However, we are free to take $j \neq \pm 1$ in $Z^*_{p^{k-l_c}}$,
therefore obtaining a contradiction, except if $Z^*_{p^{k-l_c}} =
\{1\}$ or $\{+1,-1\}$, that is if $p=3$ and $l_c=k-1$,
$p=2$ and $l_c=k-2$ or $k-1$. These are the cases recorded in the
equations (4.4$a$--$b$).

Except in the above rather special cases for $p=2$ or 3, we obtain
$l_c=l_{c+d}=k$ and so $(c,d,-c-d)$ is a permutation of
$(a,b,-a-b)$ mod $p^k$ since these six numbers are all zero.

\medskip \noindent
$\underline{\hbox{Case 4.} \quad l_a=l_{a+b}<l_b=k.}$\\
As in the case 3, the equation (4.2) shows that one of $c,d,c+d$
must be zero mod $p^k$. Again we assume $l_c=l_{c+d} \leq l_d=k$.
The equalities $l_c=l_{c+d}=l_d=k$, according to the Case 3,
are consistent with $l_a=l_{a+b}<l_b=k$ only in the
exceptional cases, {\sl i.e.} $l_a=k-2$ ($p=2$) or
$l_a=k-1$ ($p=2,3$), as shown in (4.4$c$--$d$).

We are left with $l_c=l_{c+d}<l_d=k$, so that the situation is now
symmetric with respect to the exchange of the triplets $(a,b,-a-b)$
and $(c,d,-c-d)$.
As a first step, we show that $l_c \geq l_a$.

If $l_c < l_a$, we take $\s_\a \neq 1$ in
${\rm Gal}(\z_n/\z_{n/p^{l_a}})/
{\rm Gal}(\z_n/\z_{n/p^{l_c}}) \sim Z_{p^{l_a-l_c}}$
and obtain the equation (4.11) as
before. So we have $d+(\a+1)c=0 \bmod n$, but here $\a c=c_1 p^k +
(1+j p^{k-l_a}) \g p^{l_c} {n \over p^k}$ with $j \neq 0$ in
$Z_{p^{l_a-l_c}}$. It implies $(1+\a)c = 0 \bmod p^k$ or
$$2+j p^{k-l_a} = 0 \bmod p^{k-l_c}, \qquad {\rm for\ all\ } j \neq
0 {\rm \ in\ } Z_{p^{l_a-l_c}}.
\eqno(4.13) $$
One obtains from (4.13) that $2=0 \bmod p^{k-l_a}$, or
$p^{k-l_a}=2$. From this, the equation (4.13) implies $j=-1 \bmod
p^{k-l_c-1}$ for all $j \neq 0$ in $Z_{p^{l_a-l_c}} =
Z_{p^{k-l_c-1}}$. This is a
contradiction unless $p^{k-l_c-1}=2$. Thus $p=2$, $l_a=k-1$,
$l_c=k-2$ is the only case that escapes the conclusion $l_c \geq l_a$.

We can repeat the above
argument in which we exchange the two triplets $(a,b,a+b)$ and
$(c,d,c+d)$. Doing so, we get $l_a \geq l_c$ unless $p=2$, $l_a=k-2$
and $l_c=k-1$.

Combining the two parts, we conclude that $l_a=l_{a+b}=l_c=l_{c+d}$,
except if $p=2$, $l_a=k-1$, $l_c=k-2$ or the other way round,
which are the cases listed in (4.4$e$--$f$).
For the rest, we ignore them and set $l=l_a=l_{a+b}=
l_c=l_{c+d}<k$. To complete the proof,
we still have to show that $(a,-a-b)$ is a permutation of
$(c,-c-d)$ mod $p^k$, or equivalently, that $a=\pm c \bmod p^k$.

Set $a=\a_1 p^l {n \over p^k} + \a_2 p^k$, $b=\b_2 p^k$, $c=\g_1 p^l
{n \over p^k} + \g_2 p^k$ and $d=\d_2 p^k$ with $\a_1$ and $\g_1$
coprime with $p$. Equation (4.1) in the \lq additive' form (3.7)
(with $l=1$) yields
$$\eqalign{
\z_{p^{k-l}}^{\a_1} \, & (\z_{n/p^k}^{\a_2} - \z_{n/p^k}^{\a_2+\b_2}) -
\z_{p^{k-l}}^{-\a_1} \, (\z_{n/p^k}^{-\a_2} - \z_{n/p^k}^{-\a_2-\b_2}) -
\z_{p^{k-l}}^{\g_1} \, (\z_{n/p^k}^{\g_2} - \z_{n/p^k}^{\g_2+\d_2}) \cr
& + \z_{p^{k-l}}^{-\g_1} \, (\z_{n/p^k}^{-\g_2} -
\z_{n/p^k}^{-\g_2-\d_2}) + (\z_{n/p^k}^{\b_2} - \z_{n/p^k}^{\d_2} -
\z_{n/p^k}^{-\b_2} + \z_{n/p^k}^{-\d_2}) = 0. \cr}
\eqno(4.14) $$
Note that, because $\a_1$ and $\g_1$ are coprime with $p$,
we have $\a_1 \neq -\a_1 \bmod p^{k-l}$ and $\g_1 \neq -\g_1\bmod
p^{k-l}$ unless $p^{k-l}=2$, but in this case $\a_1=\g_1=1$ from
which the claim follows since $a=c \bmod p^k$. We must show that
$\a_1 = \pm \g_1 \bmod p^{k-l}$. Suppose the contrary,
$\a_1 \neq \g_1$ {\it and} $\a_1 \neq -\g_1$. It implies that the
five powers $\z_{p^{k-l}}^{\pm \a_1},\,\z_{p^{k-l}}^{\pm \g_1}$ and 1
are all distinct. We prove that this leads to a contradiction.

If the five powers of $\z_{p^{k-l}}$ entering (4.14) are linearly
independent over $Q$, and therefore also over $Q(\z_{n/p^k})$, the
corresponding five coefficients must vanish. Setting the coefficient
of $\z_{p^{k-l}}^{\a_1}$
equal to zero leads to $\b_2=0 \bmod {n \over p^k}$, which implies
$b=0 \bmod n$, contrary to the assumption stated in the lemma.

On the other hand, if the five powers of $\z_{p^{k-l}}$ are not
independent, one of the relations (3.6$r$) must hold among them. Since
each such relation involves $p$ terms, this is impossible for $p
\geq 7$. We consider the other values of $p$ separately.

If $p=5$, the relation must be the one corresponding to $r=0$
because it is the
only one that contains 1. But then the numbers $\{\pm \a_1,\pm
\g_1\}$ must be identified with $\{5^{k-l-1},2.5^{k-l-1},3.5^{k-l-1},
4.5^{k-l-1}\}$, which is impossible since $\a_1$ and $\g_1$ are
coprime with 5, unless $k-l=1$.
If $k-l=1$, the five powers satisfy the
relation $1+\z_5+\z_5^2+\z_5^3+\z_5^4=0$. Eliminating one of them
in terms of the other (independent) ones, the equation (4.14)
implies that the five coefficients must be equal.
Making the coefficients of $\z_{p^{k-l}}^{\a_1}$ and
$\z_{p^{k-l}}^{-\a_1}$ equal, we obtain $\a_2=\pm (\a_2+\b_2) \bmod
{n \over p^k}$. The solution with the $+$ sign must be rejected as it
implies $\b_2=0 \bmod {n \over p^k}$ and $b=0 \bmod n$. Hence
$2\a_2+\b_2=0$. Repeating the argument for the coefficients of
$\z_{p^{k-l}}^{\g_1}$ and $\z_{p^{k-l}}^{-\g_1}$, we have
$2\g_2+\d_2=0$ as well. Equating now the coefficients of
$\z_{p^{k-l}}^{\a_1}$ and $\z_{p^{k-l}}^{\g_1}$, we obtain $2\a_2 =
-2\g_2 \bmod {n \over p^k}$. Finally the last condition comes from
making the coefficients of $\z_{p^{k-l}}^{\a_1}$ and 1 equal, which,
using the relations between $\b_2,\,\g_2,\,\d_2$ and $\a_2$, reads
$$\sin{2 \pi \a_2 \over n/p^k} = -4 \sin{2 \pi \a_2 \over n/p^k}
\cos{2 \pi \a_2 \over n/p^k}.
\eqno(4.15) $$
The factor $\sin{2 \pi \a_2 \over n/p^k}$ cannot be zero, because if
it was, $2\a_2$ would be zero, implying $\b_2=0$ and $b=0 \bmod n$.
Therefore the equation (4.15) reduces to $\cos{2 \pi \a_2 \over
n/p^k}=-{1 \over 4}$. The solutions of this quadratic equation read
$\z_{n/p^k}^{\a_2}=-{1 \over 4}\,(1 \pm \sqrt{-15})$, which is
impossible because $\sqrt{-15}$ does not belong to $Q(\z_{n/p^k})$
when $n \over p^k$ is coprime with 5.
More simply, $\cos{2 \pi \a_2 \over n/p^k}=-{1 \over 4}$ can be recast
into $\z_{n/p^k}^{\a_2} + \z_{n/p^k}^{-\a_2} = -\half$, expressing
a cyclotomic integer as a rational non--integer number, a plain
contradiction.

Take $p=3$. There can be a 3--term cyclotomic relation among the
five powers $\z_{p^{k-l}}^{\pm \a_1},\,\z_{p^{k-l}}^{\pm \g_1}$ and
1, but two powers will be left over. Their coefficient
must vanish, implying either $\b_2=0$ or $\d_2=0$, {\sl i.e.} $b=0$
or $d=0$ modulo $n$, a contradiction to the assumptions.

The last case is $p=2$. We assume $p^{k-l} \geq 16$ (to have
five different powers). In order to escape the conclusion $\b_2=0$
or $\d_2=0$ as for $p=3$, there must be two cyclotomic relations
among the four powers $\z_{p^{k-l}}^{\pm \a_1},\,\z_{p^{k-l}}^{\pm
\g_1}$. The coefficient of the left--over power 1 must vanish,
yielding $\b_2=\d_2$. The 2--term relation involving
$\z_{p^{k-l}}^{\a_1}$ can be
$\z_{p^{k-l}}^{\a_1}+\z_{p^{k-l}}^{-\a_1}=0$,
$\z_{p^{k-l}}^{\a_1}+\z_{p^{k-l}}^{\g_1}=0$, or
$\z_{p^{k-l}}^{\a_1}+\z_{p^{k-l}}^{-\g_1}=0$. It is easy to see that
none of them is tenable.
This finishes the proof of the lemma. \blacksquare

\bigskip
The first lemma is very restrictive and allows us to prove
the announced local version of the Theorem.

\vskip 1truecm \noindent
{\bf Corollary 1.}
Let $(c,d)=\s(a,b)$ the image of $(a,b) \in B_n$ by an automorphism.
Then $(c,d,-c-d)$ is a permutation $\pi_i$ of $(a,b,-a-b)$ mod
$p_i^{k_i}$, for any $p_i^{k_i}$ dividing $n$, or else $p=2$, and we
have, up to permutations, $a=b=c=0 \bmod 2^k$ and $d=0 \bmod
2^{k-1}$.

\noindent {\sl Proof.}
Define $m_i={n \over p_i^{k_i}}$ for $i=1, \ldots ,s$.
We first show that all the $(m_i,m_i) \in B_n$ must be left invariant
by the automorphism $\s$. Let $(c,c)=\s(m_i,m_i)$ (necessarily a
diagonal root from the discussion below the
equation (3.10)). Equation (4.1) reads
$$(1-\z_{p^k})^2 (1-\z_{p^k}^{-2}) = (1-\z_{n}^c)^2 (1-\z_{n}^{-2c}),
\eqno(4.16) $$
where, for simplicity, we dropped the index $i$ from $p_i$, $k_i$ and
$m_i$. From (3.2), the norm $\no_{Q(\z_n)/Q}$ of the left-hand side
of (4.16) is a (strictly positive) power of $p$. (The norm could be zero
if $p^k=2$, but in that case $(m,m)=({n \over 2},{n \over 2})$
is not in $B_n$.) If the
same is to be true of the right-hand side, $c$ must be a multiple
of $m \over 2$, or of $m$ if $m$ is odd, since otherwise the norm of the
right--hand side of (4.16) is either equal to 1 or equal to the power of
a prime different from $p$. In case $c={m \over 2} \bmod m$ or
equivalently
$c=\gamma m+{n \over 2}$ (hence $m$ is even and $p$ is odd), the norm
 from $Q(\z_n)$ to $Q$ of $1-\z_n^c=1+\z_{p^k}^\gamma$ is equal
to 1. Thus the norm of (4.16) requires (remember $p$ is odd)
$$\no_{Q(\z_n)/Q}\,(1-\z_n^{-2c}) = \no_{Q(\z_n)/Q} \, \big[
(1-\z_{p^k})^2 (1-\z_{p^k}^{-2}) \big] = p^{3\varphi(m)}.
\eqno(4.17) $$
Equation (4.17) has no solution for $c$ unless $p=3$ and
$\ord_3 \,\gamma=1$. (It implies $p^k \geq 9$ if $(c,c)$ is to be
in $B_n$.) In this case, the equation (4.16) can be recast into
$$(1-\z_{3^k})^2 (1-\z_{3^k}^{-2}) (1-\z_{3^k}^\g)^2 =
(1-\z_{3^k}^{2\g})^2 (1-\z_{3^k}^{-2\g}).
\eqno(4.18) $$
Then using an argument similar to that of Case 2 in Lemma 1 shows
that (4.18) has no solution for $\gamma$.
We conclude that the assumption that $c$ is not a multiple
of $m$ is contradictory.

Setting $c=\g m$, the equation (4.16) becomes
$$(1-\z_{p^k})^2 (1-\z_{p^k}^{-2}) = (1-\z_{p^k}^\g)^2
(1-\z_{p^k}^{-2\g}).
\eqno(4.19) $$
The first lemma with $n=p^k$ implies that $(\g,\g,-2\g)$ is a
permutation of $(1,1,-2)$, {\sl i.e.} $\g=1$ and $c=m$. We thus
obtain $\s(m_i,m_i)=(m_i,m_i)$ for any $m_i={n \over p_i^{k_i}}$ except
$m_i={n \over 2}$. The first step of the proof, namely $c$ must be a
multiple of $m$, can alternatively be obtained by combining the
arithmetical symmetry (3.10) (in which we take $\nu=1 \bmod p^k$)
with norm arguments. As to the second step, namely $c=\g m$ implies
$\g=1$, it also follows from the classification of simple currents [16].

\smallskip
Since all the $m_i$ are left invariant by the automorphisms,
we obtain that, for any $(a,b)$, the pairs $(a,b)$ and
$(c,d)=\s(a,b)$ must satisfy the equations (4.1) and (4.2) with $p^k$
replaced by any $p_i^{k_i} \neq 2$. Using again the first lemma with $p$
being any $p_i$, we obtain that $(c,d,-c-d)$ is a permutation of
$(a,b,-a-b)$ mod $p_i^{k_i}$, except possibly if one of the
equations (4.4) holds. Apart from the equations (4.4$a$) and (4.4$c$)
for $p=2$, we now show that the others are not compatible with (4.1).

Let us first consider the exception (4.4$a$) with $p=3$. We suppose
$a=b=c+d=0 \bmod 3^k$ and $\ord _3\,c=\ord _3\,d=k-1$. Setting $a=\a
3^k$, $b=\b 3^k$, $c=\g 3^k+{n \over 3}$ and $d=\d 3^k+{2n \over
3}$, one obtains from (4.1) with $\omega = \z_3$
$$(1-\z^\a)(1-\z^\b)(1-\z^{-\a-\b}) = (1-\omega \z^\g)
(1-\omega ^2 \z^\d)(1-\z^{-\g-\d}), \qquad \z=\z_{n/3^k}.
\eqno(4.20) $$
Expanding (4.20) in powers of $\omega$ and
setting to zero the coefficients of $\omega$ and 1 (using $1+\omega
+\omega^2=0$ to eliminate $\omega^2$) yield
respectively $\g=\d \bmod {n \over 3^k}$ and the condition
$$(1-\z^\a)(1-\z^\b)(1-\z^{-\a-\b}) = \z^\g + \z^{2\g} - \z^{-\g} -
\z^{-2\g}.
\eqno(4.21) $$
If $n \over 3^k$ is a prime power, then $(\a,\b,-\a-\b)$ is a
permutation of $(\g,\g,-2\g)$ from Lemma 1. Since the situation is still
symmetric in $(\a,\b,-\a-\b)$, we may take $\a=\b=\g$, in which case
(4.21) reduces to $\z^\g=\z^{-\g}$, contradicting $(c,d) \in B_n$.

If on the other hand, $n \over 3^k$ is not a prime power,
then there exists a prime power $q^l \mid {n \over 3^k}$ such that
$(\a,\b,-\a-\b)$ is a permutation of $(\g,\g,-2\g)$ modulo $q^l$,
that is $q \neq 2$. Furthermore we can assume $\g \neq 0
\bmod q^l$. (If $\g = 0 \bmod q^l$ for every $q \neq 2$, then $\a =
\b = 0 \bmod q^l$ as well, and we are back to (4.21) with an
effective $\z=\z_{2^{k_2}}$, a case already discussed.)
Again we choose $\a=\b=\g \bmod q^l$. Equation (4.21) reads
$$\z_{q^l}^\g(\xi^\a + \xi^\b + \xi^\g) + \z_{q^l}^{2\g}(\xi^{2\g} -
\xi^{\a+\b}) - {\rm c.c.} = 0, \qquad \xi=\z_{n/3^k q^l}.
\eqno(4.22) $$
Since $q \geq 5$ and $\g \neq 0 \bmod q^l$, the four powers of
$\z_{q^l}$ in (4.22) are linearly independent. The corresponding
coefficients must vanish, implying in particular $\xi^\a + \xi^\b
+ \xi^\g = 0$. This last equation has no solution since 3 does not
divide $n \over 3^k q^l$.

Thus the exceptions (4.4$a$) and (4.4$c$) for $p=3$ are ruled out.
Cases (4.4$b,d$--$f$) must be similarly excluded. \blacksquare

\bigskip
Note that if $n$ is a prime power, Corollary 1 is the same as
the Theorem. For composite $n$, apart from the exception for $p=2$,
all that is yet to be proved is that the
permutations $\pi_i$ in Corollary 1 cannot depend on $i$.

\vskip 0.5truecm \noindent
{\bf 5. Proof of the Theorem.}

\noindent
In order to prove that the permutations $\pi_i$ of Corollary 1 cannot
depend on $i$, we first note the following
\vskip 0.5truecm \noindent
{\bf Corollary 2.}
The diagonal roots of $B_n$ are left invariant by the automorphisms,
{\sl i.e.} $\s(a,a)=(a,a)$.

\noindent {\sl Proof.}
Since the image by $\s$ of $(a,a)$ must be a diagonal root,
we have $c=d$ in Corollary 1. If $(c,c,-2c)=\pi _i(a,a,-2a) \bmod
p_i^{k_i}$ for all $i$, the permutations $\pi_i$ can
only be the identity. So the only case to worry about is when
$(c,c,-2c)=\pi_i(a,a,-2a) \bmod p_i^{k_i}$ for $p_i \neq 2$,
yielding $c=a \bmod {n \over 2^{k_2}}$, and $a=0 \bmod 2^{k_2}$,
$c=2^{k_2-1} \bmod 2^{k_2}$ (or $a$ and $c$ interchanged). In this
case, the equation (4.1) requires $1-\z_{n/2^{k_2}}^a = \pm
(1+\z_{n/2^{k_2}}^a)$, which has no solution. \blacksquare

\bigskip
Define $\mij={n \over p_i^{k_i} p_j^{k_j}}$ for $1 \leq i \neq j
\leq s$. Since $\s(\mij,\mij)=(\mij,\mij)$,
the pairs $(c,d)=\s(a,b)$ must satisfy the new set of equations
$[S,\s]_{(\mij,\mij),(a,b)}=0$ for any $\mij \in B_n$. If, to save
the notation, one
sets $m={n \over p^k q^l}$, with $p^k \neq q^l$ any prime powers
$p_i^{k_i},\,p_j^{k_j}$ dividing $n$, these equations read
$$(1-\z_{p^k q^l}^a)(1-\z_{p^k q^l}^b)(1-\z_{p^k q^l}^{-a-b}) =
(1-\z_{p^k q^l}^c)(1-\z_{p^k q^l}^d)(1-\z_{p^k q^l}^{-c-d}).
\eqno(5.1) $$
\vskip 0.5truecm \noindent
{\bf Lemma 2.} Let $p$ and $q$ be two different primes and
$(c,d)=\s(a,b)$ two weights of $B_n$.
If $a,b,a+b \neq 0 \bmod p^k q^l$, then $(c,d,-c-d)$ is a
permutation of $(a,b,-a-b)$ mod $p^k q^l$.

\noindent {\sl Proof.} We may assume $p > q$, so that $p \geq 3$. We
also note that $a,b,a+b \neq 0 \bmod p^k q^l$ implies $c,d,c+d \neq
0 \bmod p^k q^l$ (neither side of (5.1) vanishes).
Let us define
$$\eqalign{
& a = \a_p q^l + \a_q p^k \bmod p^k q^l, \qquad
b = \b_p q^l + \b_q p^k \bmod p^k q^l, \cr
& c = \g_p q^l + \g_q p^k \bmod p^k q^l, \qquad
d = \d_p q^l + \d_q p^k \bmod p^k q^l. \cr}
\eqno(5.2) $$
 From Corollary 1, we have
$$(\g_p,\d_p,-\g_p-\d_p)=\pi_p(\a_p,\b_p,-\a_p-\b_p) \bmod p^k,
\qquad \pi_p \in S_3.
\eqno(5.3) $$
The problem being completely symmetric under a permutation of $c,d$
and $-c-d$, we fix that freedom by requiring $\pi_p=1$, so that
$\g_p=\a_p$ and $\d_p=\b_p$. We aim at proving $\pi_q=1$ as well,
{\sl i.e.} $\g_q=\a_q$ and $\d_q=\b_q$.

With $\pi_p=1$, the equation (5.1) reads
$$\eqalign{
\zp^{\a_p} (\zq^{\a_q} & - \zq^{\g_q}) - \zp^{-\a_p} (\zq^{-\a_q} -
\zq^{-\g_q}) +
\zp^{\b_p} (\zq^{\b_q} - \zq^{\d_q}) - \zp^{-\b_p} (\zq^{-\b_q} -
\zq^{-\d_q}) \cr
& - \zp^{\a_p+\b_p} (\zq^{\a_q+\b_q} - \zq^{\g_q+\d_q})
+ \zp^{-\a_p-\b_p} (\zq^{-\a_q-\b_q} - \zq^{-\g_q-\d_q}) = 0. \cr}
\eqno(5.4) $$
As often with additive equations, different cases must be
distinguished. First, there is the question as to how many among
the numbers $\a_p,\b_p,\a_p+\b_p$ are zero modulo $p^k$. There can be
zero, one or three. The easy case is when all three are zero,
because there is nothing much to prove. From Corollary 1, we have
$(\g_q,\d_q,-\g_q-\d_q)=\pi_q (\a_q,\b_q,-\a_q-\b_q)$ (the
exception for $q=2$ plays no role because of the assumption
$a,b,a+b \neq 0 \bmod p^k q^l$). Setting $\pi_p=1$ does not
fix anything (any $\pi_p$ has the same effect) and we can
harmlessly choose $\pi_p=\pi_q$ whatever $\pi_q$ is.

\medskip
Suppose one of $\a_p,\b_p,\a_p+\b_p$ is zero, $\b_p=0$ say. Then the
powers $1$ and $\z_{p^k}^{\pm \a_p}$ are all different (remember $p
\geq 3$). If $p \geq
5$ they are linearly independent, so that the corresponding three
coefficients must vanish. The coefficient of 1 being zero implies
$\d_q=\b_q$ or $\d_q={q^l \over 2}-\b_q$, while the coefficient of
$\zp^{\a_p}$ set to zero yields
$$ \zq^{\a_q} - \zq^{\g_q} = \zq^{\a_q+\b_q} - \zq^{\g_q+\d_q}.
\eqno(5.5) $$
If $\d_q=\b_q$, (5.5) obviously gives $\g_q=\a_q$. If
$\d_q={q^l \over 2}-\b_q$, the equation (5.5) becomes $\zq^{\a_q}
(1-\zq^{\b_q}) = \zq^{\g_q-\b_q} (1+\zq^{\b_q})$, so that
$(1-\zq^{\b_q})/(1+\zq^{\b_q})=\pm i$ is a purely imaginary root of
unity. In turn this means $\zq^{\b_q}=\mp i$, and again
$\g_q=\a_q$, $\d_q={q^l \over 2}-\b_q=\b_q$.

If $p=3$ ($q=2$), the powers $1$ and $\zp^{\pm \a_p}$ are again
independent,
in which case we reach the conclusion $\g_q=\a_q,\,\d_q=\b_q$, or
else $\a_p=\pm p^{k-1}$. In the latter case, equation (5.4) (with
$\b_p=0$) implies the equality of the
three coefficients
$$\eqalignno{
& \zq^{\a_q} - \zq^{\g_q} - \zq^{\a_q+\b_q} + \zq^{\g_q+\d_q} =
-\zq^{-\a_q} + \zq^{-\g_q} + \zq^{-\a_q-\b_q} - \zq^{-\g_q-\d_q},
&(5.6a) \cr
& \zq^{\a_q} - \zq^{\g_q} - \zq^{\a_q+\b_q} + \zq^{\g_q+\d_q} =
\zq^{\b_q} - \zq^{\d_q} - \zq^{-\b_q} + \zq^{-\d_q}. &(5.6b) \cr}
$$
Again from Corollary 1, $(\g_q,\d_q,-\g_q-\d_q)$ must be a
permutation $\pi_q$ of $(\a_q,\b_q,-\a_q-\b_q)$.
Trying each of the five $\pi_q \neq 1$, we
end up with impossible equations or contradictions to
$a,b,a+b \neq 0 \bmod p^k q^l$, or else $\a_q$ and $\b_q$ are related
in such a way that $\g_q=\a_q$ and $\b_q=\d_q$ still hold. Thus
$\pi_q=1$.

\smallskip
We turn to the last case: none of $\a_p,\b_p,\a_p+\b_p$ is zero. We
distinguish the cases $p \geq 5$ from $p=3$.

For $p \geq 5$, there cannot be any cyclotomic relation among the six
powers of $\zp$ entering (5.4). (For $p \geq 7$, it is obvious,
while for $p=5$, the would--be relation has to be (3.6$r$) with $r=0$
because it must contain one of the powers along with its complex
conjugate. But then one of the powers must be 1.) Therefore those
which are different are linearly independent and their coefficient
must vanish. This still leaves two possibilities: the six powers are
different or only four of them are different. The first case clearly
yields $\g_q=\a_q$ and $\d_q=\b_q$. The second possibility arises if
$\a_p=\b_p$ or $\a_p=-\a_p-\b_p$. (Any other identification
contradicts $\a_p,\b_p,\a_p+\b_p \neq 0 \bmod p^k$.) If $\a_p=\b_p$,
one obtains from (5.4) either $(\g_q,\d_q,-\g_q-\d_q)=(\a_q,\b_q,
-\a_q-\b_q)$ ({\sl i.e.} $\pi_q=1$) or
$(\g_q,\d_q,-\g_q-\d_q)=(\b_q,\a_q,-\a_q-\b_q)$, that is $\pi_q$
exchanges the first two objects, $\pi_q(1,2,3)=(2,1,3)$. But since
$\a_p=\b_p$, we could as well have fixed $\pi_p$ by requiring
$\pi_p(1,2,3)=(2,1,3)$, in which case we have $\pi_p=\pi_q$. (The
permutation $\pi_q$ is only defined relative to $\pi_p$.) The
other case with four different powers of $\zp$, namely
$\a_p=-\a_p-\b_p$, is treated similarly.

Finally we set $p=3$ and make the same kind of discussion.
First there cannot be a cyclotomic relation $\zp^x+\zp^y+\zp^z=0$,
with $x,y,z$ chosen from $\pm \a_p,\, \pm \b_p,\, \pm
(\a_p+\b_p)$. Because if there is, the triplet $(x,y,z)$ must be equal
to $(r,r+p^{k-1},r+2.p^{k-1})$ for some $r$. However every choice of
$x,y,z$ contradicts $\a_p,\b_p,\a_p+\b_p \neq 0$. Thus those powers
of $\z_{p^k}$ in (5.4) which are different must have a vanishing
coefficient. If the six powers are all different,
(5.4) gives $\g_q=\a_q$ and $\d_q=\b_q$. If they are not all
different, there are only two possibilities as in the previous case
$p \geq 5$: $\a_p=\b_p$ or $\a_p=-\a_p-\b_p$. (Here however both
equalities may hold at the same time.) We only consider the first
case, $\a_p=\b_p$, the other being similar.

If $\a_p=\b_p$ but $\a_p \neq -\a_p-\b_p$, the four powers
$\zp^{\pm \a_p}$ and $\zp^{\pm (\a_p+\b_p)}$ are different
and we obtain $\pi_p=\pi_q$ as in the $p \geq 5$ case.
If $\a_p=\b_p=-\a_p-\b_p$, the two left--over powers $\zp^{\pm \a_p}$
are different. Their coefficient must vanish, yielding the following
condition
$$
\zq^{\a_q} + \zq^{\b_q}
+ \zq^{-\a_q-\b_q} = \zq^{\g_q}
+ \zq^{\d_q} + \zq^{-\g_q-\d_q}.
\eqno(5.7) $$
If $(\g_q,\d_q,-\g_q-\d_q)=\pi_q (\a_q,\b_q,-\a_q-\b_q)$ is a
permutation, we can choose $\pi_p=\pi_q$ whatever $\pi_q$ is
(since $\a=\b=-\a-\b$). If, on the other hand, $\a_q,\b_q,\g_q,\d_q$
appear as the exception of Corollary 1, we readily check that (5.7)
is not satisfied. \blacksquare

\bigskip
We can now complete the proof. If $n$ is
composed of only two primes, the Lemma 2 proves the final result:
$(c,d,-c-d)$ is a permutation of $(a,b,-a-b)$ mod $n$. Therefore
we may assume that at least three different primes divide $n$.
Let $(c,d)=\s(a,b)$.

\medskip
Let us split the set of primes dividing $n$ into two subsets, $B$
and $G$. $B$ will contain those primes $p_i$ such $a,b,a+b$
are all 0 mod $p_i^{k_i}$, while $G$ receives the primes which
are not in $B$. Note that if $p_i$ is in $G$, then at most one among
$a,b,a+b$ can be zero modulo $p_i^{k_i}$, and we accordingly split $G$
into four subsets:
$$\eqalign{
& G_0 = \{p_i \in G \;:\; a,b,a+b \neq 0 \bmod p_i^{k_i} \}, \cr
& G_1 = \{p_i \in G \;:\; (a,b,-a-b)=(0,b,-b) \bmod p_i^{k_i}
\}, \cr
& G_2 = \{p_i \in G \;:\; (a,b,-a-b)=(a,0,-a) \bmod p_i^{k_i}
\}, \cr
& G_3 = \{p_i \in G \;:\; (a,b,-a-b)=(a,-a,0) \bmod p_i^{k_i}
\}. \cr}
\eqno(5.8) $$
We first prove that $(c,d,-c-d)$ is a permutation of $(a,b,-a-b)$
modulo $G$, and by this we mean modulo $\prod_{p_i \in G}\,p_i^{k_i}$.

If $G_0 \neq \emptyset$, it contains a prime $p_1$ such that
$a,b,a+b \neq 0 \bmod p_1^{k_1} p_i^{k_i}$ for every $p_i \neq p_1$.
Then the Lemma 2 implies that $(c,d,-c-d)$ is a permutation of
$(a,b,-a-b)$ mod $p_1^{k_1} p_i^{k_i}$ for all $i \geq 2$,
 from which the stronger claim clearly follows: $(c,d,-c-d)$ is a
permutation of $(a,b,-a-b)$ modulo $n$, since $\pi_i=\pi_1$ for
all $i \geq 2$.

If $G_0 = \emptyset$, at least two of the subsets $G_1,G_2,G_3$ are
non--empty, since
otherwise it would contradict $a,b,a+b \neq 0 \bmod n$. From the
definitions (5.8), it follows that if $p_i$ and $p_j$ belong to two
different subsets $G_k$, then $a,b,a+b \neq 0 \bmod p_i^{k_i}
p_j^{k_j}$ and $(c,d,-c-d)$ is a permutation of $(a,b,-a-b) \bmod
p_i^{k_i} p_j^{k_j}$. By making $i$ and $j$ vary over the three
subsets (but keeping $p_i$ and $p_j$ in different $G_k$), we obtain
the same result for any pair $p_i,\,
p_j$ of primes in $G$, whether in different subsets or not. Again the
statement follows: $(c,d,-c-d)$ is a permutation of $(a,b,-a-b)$ mod $G$.

\smallskip
We now consider the primes in $B$. For the primes $p_i$ in $B$
different from 2, we know that $(c,d,-c-d)$ is permutation of
$(a,b,-a-b)=(0,0,0) \bmod p_i^{k_i}$. Which permutation it is
becomes irrelevant since the three objects are identical anyway.
We can therefore choose the same permutation as the one relating
$(c,d,-c-d)$ to $(a,b,-a-b) \bmod G$, and doing so we obtain
$$(c,d,-c-d) = \pi (a,b,-a-b) \bmod {n \over 2^{k_2}}.
\eqno(5.9) $$
The only remaining case is when 2 is in $B$, that is when $a$ and
$b$ are both multiple of $2^{k_2}$. In this case, the
Corollary 1 does not guarantee that $c$ and $d$ are also multiple of
$2^{k_2}$. If they are, then of course the statement
(5.9) is also true mod $n$. Thus it remains to rule out the single
exception of Corollary 1, namely $a=b=0 \bmod 2^{k_2}$, and say $c=0
\bmod 2^{k_2}$, $d=2^{k_2-1} \bmod 2^{k_2}$. We can do so by
repeating the above argument in which we exchange $(a,b,-a-b)$ with
$(c,d,-c-d)$. We define two new sets $B'$ and $G'$ as above but
relative to $(c,d,-c-d)$. From $d=2^{k_2-1} \bmod 2^{k_2}$, we find
that $p_i=2$ belongs to $G'$, and since Corollary 1 and Lemma 2 are
symmetric under the interchange of $(a,b,-a-b)$ and $(c,d,-c-d)$, we
conclude that (5.9) holds modulo $n$. The proof of the Theorem is
complete.

\vskip 0.5truecm \noindent
{\bf 6. Perspectives.}

\noindent
The proof we gave for $SU(3)$ in Sections 4 and 5 has clearly a
multiplicative and an additive part. They both can be applied to any
other algebra, since in most instances, the problem is to assess
independence properties of cyclotomic numbers.
As this usually involves discussing different cases separately,
it can become rather painful when the number of terms increases.
This is especially true when additive relations must be examined. So
for practical feasability, solving the problem for large algebras
requires a more systematic way of dealing with the
additive part. Another possibility is to keep the whole discussion
at the multiplicative level, which is more satisfactory and easier
to handle, even when the number of terms gets large. Essentially,
this means changing the arguments of Lemma 2 so as to keep the
multiplicative character of equation (5.1). It would not
yield a simpler proof for $SU(3)$, but it looks more promising for
larger algebras.

\vskip 1truecm
I would like to thank E. Thiran and J. Weyers for discussions during
the early stage of this work.
While this manuscript was being completed, I received the preliminary
version of a preprint by T. Gannon which contains the
full classification of $SU(3)$ modular invariant partition functions
[17]. In particular, it also contains for $SU(3)$, by using the
results of [11], the classification of automorphisms shown here.

\vfill \eject
\parskip=5pt
\vskip 0.8truecm \noindent
{\bf References.}

\item{[1]} J. Cardy, {\it Nucl. Phys.} B270 (1986) 186.
\item{[2]} G. Moore and N. Seiberg, {\it Nucl. Phys.} B313
(1989) 16.
\item{[3]} E. Verlinde, {\it Nucl. Phys.} B300 [FS22] (1988) 360.
\item{[4]} A. Capelli, C. Itzykson and J.B. Zuber, {\it Commun.
Math. Phys.} 113 (1987) 1.
\item{[5]} A. Capelli, {\it Phys. Lett.} 185B (1987) 82.
\item{[6]} D. Gepner and Z.Qiu, {\it Nucl. Phys.} B285 (1987) 423.
\item{[7]} P. Ginsparg, {\it Nucl. Phys.} B295 (1988) 153,\\
E. Kiritsis, {\it Phys. Lett.} 217B (1989) 427.
\item{[8]} Ph. Ruelle, E. Thiran and J. Weyers, {\it Commun. Math.
Phys.} 133 (1990) 305.
\item{[9]} Ph. Ruelle, E. Thiran and J. Weyers, preprint
DIAS--STP--92--26, November 1992.
\item{[10]} V. Kac, {\it Infinite Dimensional Lie Algebras},
Birkh\"auser, Boston 1983.
\item{[11]} L. B\'egin, P. Mathieu and M. Walton, {\it Mod. Phys.
Lett.} A7 (1992) 3255.
\item{[12]} D. Bernard, {\it Nucl. Phys.} B288 (1987) 628.
\item{[13]} D. Altsch\"uler, J. Lacki and P. Zaugg, {\it Phys. Lett.}
205B (1988) 281.
\item{[14]} D. Verstegen, {\it Nucl. Phys.} B346 (1990) 349.
\item{[15]} L. Washington, {\it Introduction to Cyclotomic Fields},
GTM 83, Springer, New York 1982.
\item{[16]} J. Fuchs, {\it Commun. Math. Phys.} 136 (1991) 345.
\item{[17]} T. Gannon, {\it The Classification of Affine SU(3) Modular
Invariant Partition Functions}, December 1992.

\end